\documentclass{mem-astro-ph}
\usepackage{natbib}\usepackage{txfonts}  
\usepackage{graphicx}
\usepackage[a4paper]{hyperref}
\idline{76}{000}
\begin{document}
\def\teff{$T\rm_{eff }$}
\def\kms{$\mathrm {km s}^{-1}$}

\title{Modelling NIR molecular lines for Miras}

\author{
W. \,Nowotny\inst{1}, 
S. \, H\"ofner\inst{2},
T. \, Lebzelter\inst{1},
B. \, Aringer\inst{1}
\and J. \, Hron\inst{1}
}

 
\institute{
Institut f\"ur Astronomie der Universit\"at Wien, Vienna, Austria\\
\email{nowotny@astro.univie.ac.at}
\and
Departement of Astronomy and Space Physics, Uppsala University, 
Uppsala, Sweden
}

\authorrunning{Nowotny et al.}
\titlerunning{Modelling NIR molecular lines for Miras}

\abstract{
The atmospheric structure of Mira variables is considerably influenced by
pulsation. Molecular absorption lines in the near-infrared (NIR), 
especially second overtone CO lines, show therefore a characteristic 
behaviour in time-series of high-resolution spectra. We computed synthetic 
CO line profiles based on a new dynamic model atmosphere and derived radial 
velocities (RVs) from the Doppler shifted lines. For the first time, we 
could quantitatively reproduce observations of the very typical, 
discontinuous RV curves.

}
\maketitle{}


\section{Probing atmospheric kinematics}

Mira variables are luminous late-type giants, representing stars of low- to 
intermediate mass in the late evolutionary stage called Asymptotic Giant 
Branch (AGB). Due to the low effective temperature ($<$3500\,K), molecules
can form efficiently in the outer layers and dominate their spectral 
appearance. Towards the end of their evolution, such stars become unstable 
against radial pulsations. Eventually, this leads to the condensation of 
dust grains in the cool atmospheres and stellar winds develop.

High-resolution spectroscopy in the NIR has proven to be a valuable tool to 
investigate atmospheric kinematics. RVs derived from the wavelength shifts  
of spectral lines provide clues on the gas velocities throughout the extended, 
dynamic atmosphere. In the past, CO lines of the $\Delta\,v$=3 vibration--rotation 
bands at $\lambda$$\approx$1.6\,$\mu$m have been used to study deep photospheric 
layers in Miras, which are heavily affected by the pulsation. As the emerging 
shock front passes through the region of line formation, it leads to a very 
characteristic behaviour of the CO line profiles (line doubling around phases of 
maximum luminosity) and results in discontinuous, S-shaped RV curves during the 
lightcycle (e.g. Hinkle et al. 1984). 

\begin{figure*}[t!]
\resizebox{\hsize}{!}{\includegraphics[clip=true]{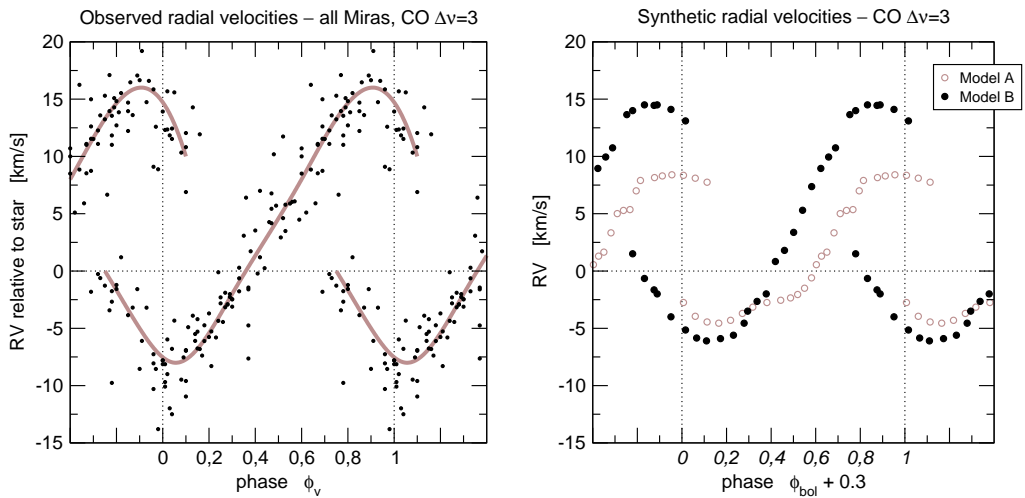}}
\caption{\footnotesize
\textit{Left:} RV curves of second overtone CO lines for a sample of typical Miras 
(data taken from Lebzelter \& Hinkle 2002). The grey, thick line was fit to guide 
the eye.
\textit{Right:} RV curves for the same lines as derived from model calculations 
(R=70.000). Model\,A was used in Nowotny et al. (2005a,b,c), while Model\,B is 
an improved version, which yields more realistic results, especially a comparable 
velocity amplitude of $\Delta$RV=20.6\,km$\cdot$s$^{-1}$.}
\label{f:rvcurves}
\end{figure*}

In Fig.\,\ref{f:rvcurves} (left), a compilation of observational data (extracted 
from FTS spectra) is shown, suggesting a similar picture for all Miras, 
independent of their parameters (spectral type, period, metallicity, etc.). 
Realistic atmospheric models should reproduce this.
%


\section{Modelling line profiles and synthetic radial velocities}

Dynamic model atmospheres are needed to reproduce the complex phenomena within
the outer layers of Mira variables. Recently, we presented synthetic line 
profiles based on models from H\"ofner et al. (2003). A special radiative 
transfer code was applied, which also takes velocity effects into account. 
Results can be found in Nowotny et al. (2005a,b,c). It could be shown, that 
state-of-the-art models can qualitatively reproduce the behaviour (line 
profiles, RVs) of spectral lines originating at different depths. 

Still, some differences remained compared to observations. For CO 
$\Delta\,v$=3 lines the line splitting appears too weak and the extracted 
velocity amplitude $\Delta$RV too low (cf. Sect.\,6.2. in Nowotny et al. 2005c).
Here, calculations are presented using a new dynamic atmospheric model, for 
which the parameters\footnote{$L_*$=7$\cdot$10$^3$L$_{\odot}$, 
$M_*$=1.5M$_{\odot}$, $T_*$=2600\,K, C/O=1.4, P=490\,d, 
$\Delta u_{\rm p}$=6\,km$\cdot$s$^{-1}$} were somewhat tuned to fit this 
aspect. Line profiles for the CO 5--2 P30 line were computed for various 
phases during a pulsation period. The spectral synthesis was carried out 
exactly as described in detail in Nowotny et al. (2005b,c). As also found 
there, the RV curves look more complicated for the highest resolution calculated 
(300.000), but become similar to observed ones if the spectra are rebinned 
to a resolution  of 70.000 (comparable to FTS spectra). Lacking visual phases 
for the models, the synthetic RVs in Fig.\,\ref{f:rvcurves} (right) are shifted 
by an arbitrary value of 0.3. It can be recognised, that the results of the 
computations resemble the observations reasonably well (line-doubling, S-shape, 
amplitude $\Delta$RV, etc.). This fundamental characteristic of Miras can now be 
reproduced even quantitatively in a self-consistent way by the new model.


\begin{acknowledgements}
This work was supported by the ``Fonds zur F\"orderung der 
Wis\-sen\-schaft\-li\-chen For\-schung'' under project number P14365--PHY 
and the Swedish Research Council.
\end{acknowledgements}

\bibliographystyle{aa}

\end{document}